\documentclass[aps,prb,superscriptaddress,showpacs,twocolumn]{revtex4}

\usepackage{amsmath,amsfonts,amssymb,graphics,graphicx,epsfig,color,times,indentfirst,layout}
\begin{document}

\title{Landau-Zener-St\"uckelberg interference in a multi-anticrossing system}

\author{ Jindan Chen }

\affiliation{National Laboratory of Solid State Microstructures and Department of
Physics, Nanjing University, Nanjing 210093, China }

\author{ Xueda Wen }
\altaffiliation[The author is now at Department of Physics, UIUC, Urbana, IL 61801, USA]{.} 
\affiliation{National Laboratory of Solid State Microstructures and Department of
Physics, Nanjing University, Nanjing 210093, China }

\author{Guozhu Sun}
\affiliation{Research Institute of Superconductor Electronics, Department of
Electronic Science and Engineering, Nanjing University, Nanjing 210093, China}

\author{ Yang Yu}
\email{ yuyang@nju.edu.cn}\affiliation{National Laboratory of Solid State Microstructures and
Department of Physics, Nanjing University, Nanjing 210093, China }
\date{\today}
\begin{abstract}

We propose a universal analytical method to study the dynamics of a multi-anticrossing system subject to driving by one single large-amplitude triangle pulse, within its time scales smaller than the dephasing time. Our approach can explain the main features of the Landau-Zener-St\"uckelberg interference patterns recently observed in a tripartite system [Nature Communications 1:51 (2010)]. In particular, we focus on the effects of the size of anticrossings on interference and compare the calculated interference patterns with numerical simulations. In addition, Fourier transform of the patterns can extract information on the energy level spectrum.

\end{abstract}

\maketitle

\section{INTRODUCTION}
Landau-Zener (LZ) transitions\cite{Zener,Oliver1,Shevchenko} at the anticrossings play a fundamental role in coherent quantum control which is key to the realization of quantum computation\cite{Mooij}. To date, great efforts have been devoted to investigate the coherent quantum dynamics of the states at energy-level crossings. In a strongly harmonic-driven two-level system (TLS), repeated LZ transitions give rise to St\"uckelberg or Ramsey-type oscillations, in analogy to Mach-Zehnder (MZ) interferometer. The MZ-type interferometry has been observed in a driven superconducting flux qubit\cite{Oliver}, a Cooper-Pair box\cite{Sillanpaa}, and a quantum dot system\cite{Petta}. These patterns have been theoretically reconstructed from different perspectives\cite{Valenzuela,Berns1,Rudner,Wei}.

In addition to Landau-Zener-St\"uckelberg (LZS) interference of one-anticrossing type, several experiments have also reported LZS interference in the multi-anticrossing level structure. In the presence of much stronger harmonic excitation, the qubit state can be driven through more of the constituent energy-level anticrossings, and the resulting LZS interference reveals complex checkerboardlike patterns\cite{Berns}. This has been explained well in theory\cite{Xueda,Ferron}. Most recently, a new method of coherent manipulation of quantum states in a tripartite quantum system formed by a superconducting qubit coupled to two TLSs is reported\cite{Guozhu}. The manipulation relies on the LZS interference produced by transitions at the two anticrossings, and has potential application in precise control of quantum states in the tripartite system. Nevertheless, a universal model to explain the observation has not been proposed. We shall show below that it can be understood rather easily using our approach.

In this work, we start with a strongly driven one-anticrossing system where we consider LZ transition as a gate operation. We then analyze the dynamics of multi-anticrossing system. Under the strong triangle pulse driving, occupation probability of the system at the initial state exhibits, as a function of the driving amplitude and the pulse width, diverse LZS interference patterns. Our approach presents a unified analytical treatment of these diverse patterns. In a specific case of two-anticrossing system, we focus on how the interference patterns are influenced by the sweep rate and the coupling strength, and thereby elucidate the underlying physics of the various patterns. Converting the patterns to the phase domain, Fourier transform of the resulting population oscillation reveals Fourier components of the compound pattern, which is in agreement with our analysis. In all the systems under study, the influence of relaxation and dephasing is neglected to obtain a clear physics picture of the underlying quantum physics.

This paper is organized as follows. In Sec.II we introduce the analytical expression to describe the dynamics for the multi-anticrossing system. In Sec.III we apply the general result to the $N=2$ case, and focus on the effect of the size of anticrossings on the interference patterns. The calculated patterns of four representative combinations of two anticrossings are in agreement with the numerical results. A discussion of the formation of the interesting dark state--one special case in the two-anticrossing system, is also included. In Sec.IV, Fourier transform of these patterns are presented, exhibiting an explicitly ordered structure of one dimensional arcs that offers energy spectrum information. Finally, Sec.V contains a summary of this work.

\section{ANALYTICAL MODEL}
\begin{figure}
\includegraphics[width=2.2in]{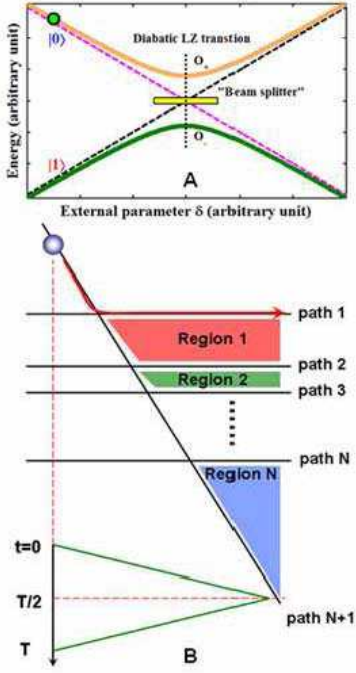}
\caption{(Color on line) Schematic diagram of (A) one anticrossing and (B) LZS interference in a phase qubit coupled to N TLSs. (A) LZ transition at the anticrossing splits the initial state into a superposition of two states in a like manner of a beam splitter. (B) Starting at the dot maker, the qubit state is swept by a triangle pulse. A succession of LZ transitions at the chain of anticrossings result in a superposition state. The state accumulates phases (shaded region), interferes at the return LZ transitions successively, and returns to the initial state with a finite probability. } \label{Anticrossing}
\end{figure}

First, we consider one-anticrossing system in which two energy levels of a quantum TLS ``cross'' each other as some external parameter is varied. At the energy-level crossing, hybridization of the two states results in an anticrossing due to coupling of the states, as shown in Fig.\ref{Anticrossing}(A). The Hamiltonian of TLS is,

\begin{equation}
H_{TLS}
=\left[\begin{array}{cccc}
\epsilon_0(t) & \Delta\\
\Delta        & \epsilon_1(t)
\end{array}\right],
\end{equation}
where $\epsilon_0(t)$ and $\epsilon_1(t)$ are the energy levels of two diabatic states, and $\Delta$ is their coupling strength. The transition between energy levels at the anticrossing is what we call LZ transition.

Following Damski and Zurek's adiabatic-impulse approximation model\cite{Shevchenko,Ashhab}, we can obtain a convenient description of the system's dynamics. It is provided that the system evolves adiabatically everywhere except at the points of minimum energy splitting where a sudden mixing in the population of the two energy levels occurs. This non-adiabatic transition at the anticrossing can be described by a unitary transformation\cite{Shevchenko,Sillanpaa} .
\begin{equation}\label{LZGate}
\hat{U}_1=\left[\begin{array}{cccc}
\cos(\theta /2)\exp(-i\tilde{\phi}_S) & i\sin(\theta /2)\\
i\sin(\theta /2)                      & \cos(\theta /2)\exp(i\tilde{\phi}_S)
\end{array}\right],
\end{equation}
where $\sin^2(\theta /2)=P_{LZ}$, where $P_{LZ}$ is the Landau-Zener transition probability at the anticrossing. If the anticrossing is swept from the infinity on one side to the infinity on the other, $P_{LZ}$ has the asymptotic form:
\begin{equation}\label{PLZ}
P_{LZ}=\exp(-2\pi\frac{\Delta^2}{\hbar\nu}),
\end{equation}
in which $\nu=d(\epsilon_1-\epsilon_0)/dt$ is the variation rate of the energy separation between the two diabatic levels, and $2\Delta$ is the size of the anticrossing. In addition, phase jump $\tilde{\phi}_S=\phi_S-\pi/2$, related to the general Stokes phenomenon\cite{Shevchenko}.$\phi_S$ is the so-called Stokes phase that takes the form:
\begin{equation}
\phi_S=\frac{\pi}{4}+\delta(\ln\delta -1)+\arg\Gamma(1-i\delta),
\end{equation}
where $\delta=\Delta ^2/\hbar\nu$ is called adiabatic parameter and $\Gamma$ is the gamma function. In the adiabatic limit $\phi_S\rightarrow 0$, and in the sudden limit $\phi_S\rightarrow \pi /4$.

From the perspective of optics, the avoided level crossing, when driven through, can be viewed as a beam splitter, because LZ transition taking place at the anticrossing splits an input state in a superposition of two states, just analogous to an optical beam splitter which splits the incident light in two. In this sense, we define reflection coefficient $|r|^2=1-P_{LZ}$ and transmission coefficient $|t|^2=P_{LZ}$.

Then, we take into account the multi-anticrossing system which can be realized in a superconducting phase qubit with many TLSs inside its Josephson Junction. A phase qubit\cite{Makhlin,You,Clarke,Martinis1} consists of a single current-biased Josephson junction. When biased close to the critical current $I_0$, the qubit can be treated as a tunable artificial atom with discrete energy levels that exist in a potential energy landscape determined by the circuit design parameters and bias. In the qubit-TLS coupled system, TLS\cite{Simmonds,Martinis,Zagoskin,Matthew} is formed in the disordered barrier material, where some atoms can occupy two positions, corresponding to two quantum states\cite{Matthew}. When the energy separation of qubit states equals that of TLS, resonant tunneling between the states opens an avoided level crossing in the energy spectrum of qubit, and forms a multi-anticrossing chain (see Fig.\ref{Anticrossing}(B)). Its Hamiltonian takes the form,
\begin{equation}
H_{qubit-NTLSs}
=\left[\begin{array}{ccccc}
\epsilon(t)      & \Delta_1   & \Delta_2   & \cdots & \Delta_N\\
\Delta_1         & \epsilon_1 & 0          & \cdots & 0\\
\Delta_2         & 0          & \epsilon_2 & \cdots & 0\\
\vdots           & \vdots     & \vdots     & \ddots & \vdots\\
\Delta_N         & 0          & 0          & \cdots & \epsilon_N
\end{array}\right].
\end{equation}
The time dependent $\epsilon(t)$ is energy spacing between the ground state and the excited state of qubit and $\epsilon_i, i=1,\cdots,N$ is energy spacing of the $i^{th}$ TLS. It is assumed that qubit is initially prepared in its excited state, and quantum state transitions are then driven using a triangle pulse with amplitude V and period T. This is a double-passage process during which anticrossing regions are passed twice. The first excursion through the anticrossings coherently divides the signal into N output paths, where the dynamical phase is accumulated during the adiabatic parts of the evolution. Then the second excursion recombines the separated signals via LZ transition, and results in the LZS interference patterns.

\begin{figure*}
\centering
\includegraphics[width=6.5in]{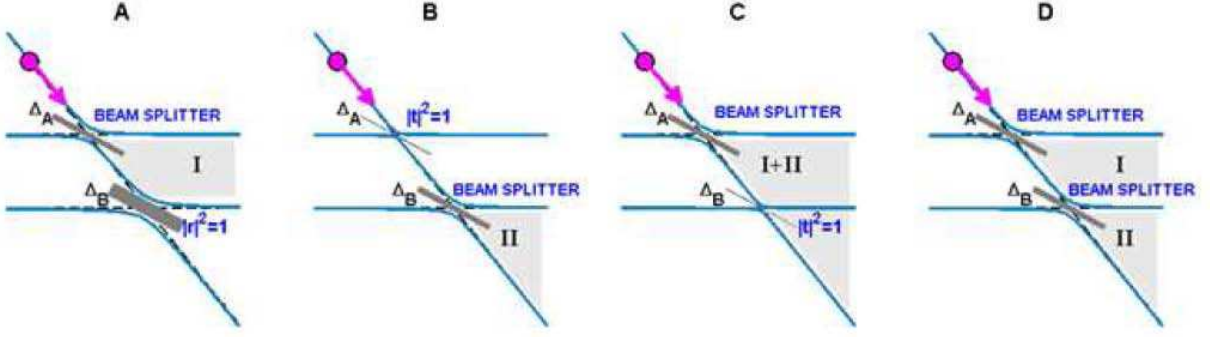}
\caption{(Color online) Different combinations of two beam-splitters. (A)$\Delta_A<\Delta_B$. M$_1$ acts as a beam splitter, while M$_2$ reflects the state with $P_{LZ}$ approaching zero. Thus the interfering pattern is mainly dependent on the phase accumulated in region I. (B)We still set $\Delta_A<\Delta_B$, but both of them are much smaller so that M$_1$ transmits the state with $P_{LZ}$ approaching unity, while M$_2$ plays the role of a beam splitter. The phase accumulated in region II plays a major role in forming the interference fringes. (C)$\Delta_A>\Delta_B$. M$_1$ acts as a beam splitter, while M$_2$ transmits the state. The phase accumulated in region I+II plays the leading part. (D)$\Delta_A=\Delta_B$, Both M$_1$ and M$_2$ act as beam splitters. $\Phi_I, \Phi_{II}, (\Phi_I+\Phi_{II})$ altogether contribute to the interference pattern. } \label{TwoAnticrossings}
\end{figure*}

The algebra describing this one full driving cycle of triangle pulse is as follows. As the anticrossings are traversed from left to right in turn, according to Eq.(\ref{LZGate}) the transition amplitude to each output path is:
\begin{equation}\label{Amplitude}
A_{out}=
\left\{
\begin{array}{ll}
\cos\theta_1 &\textrm{to path $1$}\\
\sin\theta_1\cos\theta_2                       &\textrm{to path $2$}\\
\sin\theta_1\sin\theta_2\cos\theta_3           &\textrm{to path $3$}\\
\ \ \ \ \ \vdots                               &\textrm{\ \ \ \vdots}\\
(\prod_{k=1}^{k=i-1}\sin\theta_k )\cos\theta_i &\textrm{to path $i$}\\
\ \ \ \ \ \vdots                               &\textrm{\ \ \ \vdots}\\
(\prod_{k=1}^{k=N-1}\sin\theta_k )\cos\theta_N &\textrm{to path $N$}\\
\prod_{k=1}^{k=N}\sin\theta_k                  &\textrm{to path $N+1$}
\end{array}
\right.
\end{equation}
The adiabatic evolution is in the form of $\exp(i\phi_i)$, where $\phi_i$ denotes the phase accumulated on path $i$ relative to the ground state $|0\rangle$ between the two successive crossings. When the anticrossings are traversed from right to left, the transition amplitude back to the initial state from each path $A_{in}$ equals $A_{out}$. Then after one round of LZS interference, the amplitude for the coupling system remaining in the initial state is:
\begin{equation}\label{TotalAmpli}
\begin{split}
A_{total}&=\sum_{i=path 1 }^{path N+1 }A_{out}(i)A_{in}(i)\exp(i\tilde\phi_{i})\nonumber\\
&=\sum_{i=path 1 }^{path N+1 }A_{out}^2(i)\exp(i\tilde\phi_{i}).
\end{split}
\end{equation}
Therefore, the probability for the upper level after one period of triangle pulse is given by:
\begin{equation}
\begin{split}
P_{1}&=|A_{total}|^{2}\nonumber\\
&=\sum_{i=1}^{N+1}\sum_{j=1}^{N+1}A_{out}^2(i)A_{out}^2(j)\exp(i\tilde\phi_{ij})\nonumber.
\end{split}
\end{equation}
It can be further simplified to the form,
\begin{equation}\label{Probability}
\begin{split}
P_{1}&=\sum_{i=1}^{i=N+1}A_{out}^2(i)\nonumber\\
&+2\sum_{i=1}^{N+1}\sum_{j=1}^{j<i}A_{out}^2(i)A_{out}^2(j)\cos\tilde\phi_{ij},
\end{split}
\end{equation}
where
\begin{equation}
\tilde\phi_{ij}=\tilde\phi_i-\tilde\phi_j=\sum_{n=j-1}^{n=i-1}\Phi_n. (j<i)
\end{equation}
Here $\Phi_n=\int_{t_n}^{t_n'}[E_{pathn}(t)-E_{pathn+1}(t)]dt$ (at $t=t_n (t_n')$, the $n$th anticrossing is traversed from left (right) to right (left)) is the interference phase accumulated in the area of region $n$ as shown in Fig.\ref{Anticrossing}(B).

Eq.(\ref{Probability}) indicates that the resultant interference pattern is subject to the LZ transition amplitude and the interference phase. Every two paths in the energy diagram accumulate one interference phase and give rise to one type of interference fringes (see Fig.\ref{Anticrossing}(B)). The weightings of the $C_{N+1}^2$ interference patterns governed by a single phase difference on the resultant interference depend on LZ transition amplitude. The two factors to determine the LZ transition probability are the size of the anticrossing and the velocity with which it is traversed. Therefore by varying sweep rate and coupling strength between qubit and TLSs, we can generate a variety of interference patterns with promising use in quantum control.

For example, (1) $\sin\theta_k\simeq0$, which means the size of the $kth$ anticrossing is so large that no ingredient of the wavefunction can transmit through it. Therefore, for $i>k$, $A_{out}(i)=0$,
and the number of interferences reduces from $C_{N+1}^2$ to $C_{k}^2$. Especially, if $k=2$, there is totally $C_{2}^2=1$
interference, i.e., the interference between path 1 and path 2. (2) $\cos\theta_k\simeq0$, which means the size of the $kth$
anticrossing is so small that the wavefunction cannot feel its existence. In this case, $A_{out}(k)=0$, and the $kth$ path
will not participate in the interference. For more interesting concrete examples, let's turn to Sec.III.

\begin{figure*}
\centering
\includegraphics[width=5.5in]{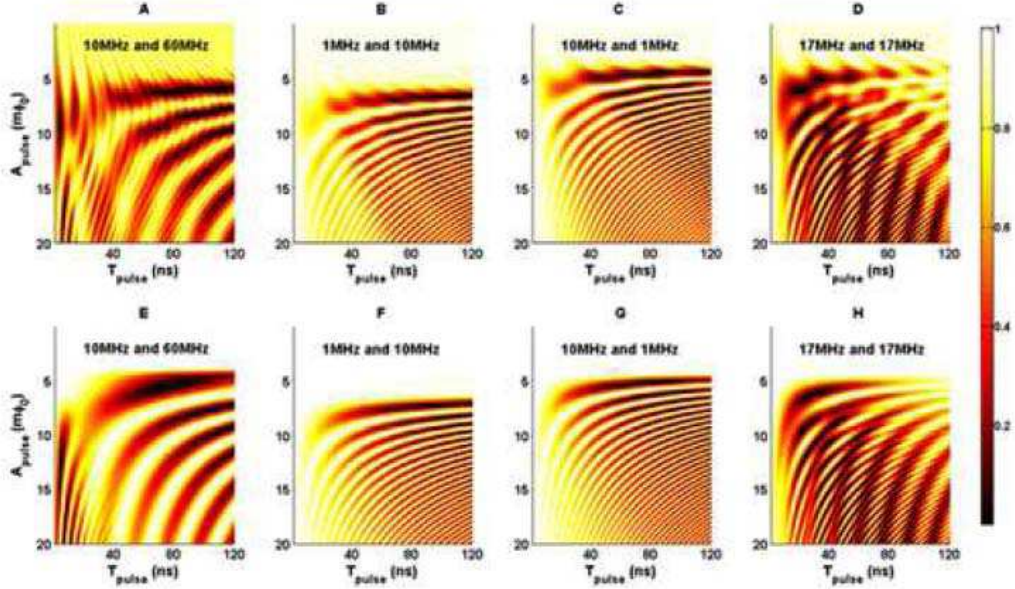}
\caption{(Color on line) (A)-(D) are numerical simulations of qubit population at the initial state$|1\rangle$, plotted as a function of the amplitude of the driving triangle pulse $A_{pulse}$ and its time width $T_{pulse}$. (E)-(H) are analytical results using Eq.(\ref{TwoTLSs}). The parameters used are extracted from our pertinent experiment \cite{Guozhu}. (A $\&$ E)$\Delta_A=10MHz, \Delta_B=60MHz$. (B $\&$ F)$\Delta_A=1MHz, \Delta_B=10MHz$. (C $\&$ G)$\Delta_A=10MHz, \Delta_B=1MHz$. (D $\&$ H)$\Delta_A=\Delta_B=17MHz$.} \label{DiffCombination}
\end{figure*}

\begin{figure*}
\centering
\includegraphics[width=5.5in]{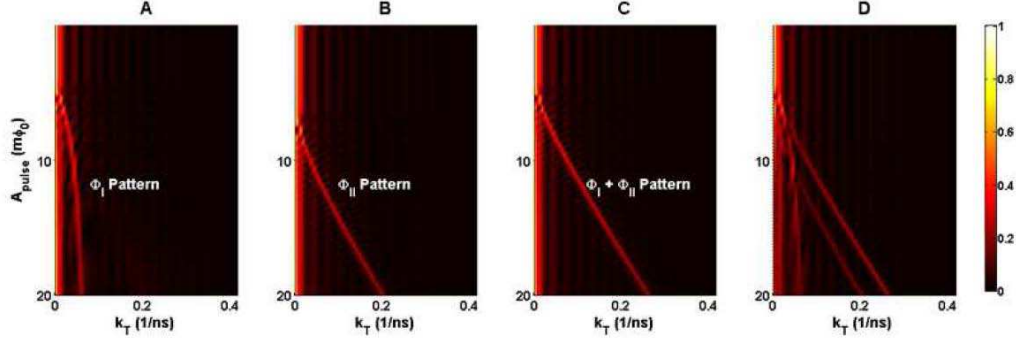}
\caption{(Color on line) Discrete Fourier Transform of the LZS patterns in Fig.\ref{DiffCombination}.
 In FT, one curve corresponds to one interference pattern. The number of the curves indicate the number of anticrossings that act as beam splitters under the pulse driving. The curves in (A) (B) (C) indicate $\Phi_I$, $\Phi_{II}$, $\Phi_I+\Phi_{II}$ pattern respectively. (D) shows all the three curves as in this case all the patterns make comparable contributions to the interference.} \label{FT}
\end{figure*}

\section{APPLICATION TO TWO ANTICROSSINGS}

LZS interference in a phase qubit coupled to two TLSs has recently been observed\cite{Guozhu}. 3D view of the probability for the initial state of qubit under one round of strong triangle pulse drive explicitly characterizes the sweep-rate-dependency of LZS interference.  This feature can be explained using the above result applied to N=2 case. In this case, the coupled Hamiltonian is,
\begin{equation}
H_{qubit-2TLSs}=\left[\begin{array}{ccccc}
\epsilon(t) & \Delta_1   & \Delta_2  \\
\Delta_1         & \epsilon_1 & 0         \\
\Delta_2         & 0          & \epsilon_2\\
\end{array}\right].
\end{equation}
According to the general formula in Eq.(\ref{Probability}), the occupation probability at the initial state $|1g_1g_2\rangle$ after one pulse driving takes the form:
\begin{eqnarray}\label{TwoTLSs}
\lefteqn{P_1=\cos^4\theta_1+\sin^4\theta_1\cos^4\theta_2+\sin^4\theta_1\sin^4\theta_2{}}\nonumber\\
&&{}+2\sin^2\theta_1\cos^2\theta_2\cos^2\theta_1\cos \Phi_I\nonumber\\
&&{}+2\sin^4\theta_1\sin^2\theta_2\cos^2\theta_2\cos \Phi_{II}\nonumber\\
&&{}+2\sin^2\theta_1\sin^2\theta_2\cos^2\theta_1\cos (\Phi_I +\Phi_{II}),
\end{eqnarray}
where $\Phi_i$(i=I,II) is the total phase accumulated in region $i$ as shown in Fig.2. It clearly exhibits that the interference fringes comprise of 3 ($C_3^2$) patterns, governed by phase accumulation in area I, II, and I+II, respectively. With the size of the two anticrossings fixed, as the sweep rate is varied, the interference pattern can be divided into three regions displaying the features of the three phase patterns respectively. In the slow limit, TLS$_1$ acts as a beam splitter and TLS$_2$ acts as a total reflection mirror; $\Phi_I$ pattern dominates. In the fast limit, TLS$_1$ acts as a total transmission mirror and TLS$_2$ acts as a beam splitter; $\Phi_{II}$ pattern makes the main contribution. In the intermediate region, both TLS$_1$ and TLS$_2$ act as beam splitters; $\Phi_I+\Phi_{II}$ pattern is the principal feature.

On the other hand, the size of anticrossing also manipulates weights of the three patterns in the resulting compound interference fringes. Though experimental realization of coupling strength control has not been achieved, investigation on the effect of the size of anticrossings on the interference patterns could make predictions of various interference fringes with potential use in future coherent control of hybrid qubit system. Here we choose four representative combinations of two TLSs, and the parameters used are extracted from G.Z.Sun's experiment\cite{Guozhu}. Based on this we discuss the effect of $\Delta$:

(i) $\Delta_A=10MHz, \Delta_B=100MHz$. In this case, $P_{LZ2}\sim 0$, $\sin\theta_2\sim 0$, $\cos\theta_2\sim 1$. The occupation probability approximately equals to:
\begin{equation}
P_1=\sin^4\theta_1+\cos^4\theta_1+2\sin^2\theta_1\cos^2\theta_1\cos \Phi_I.
\end{equation}
This combination reveals the main feature of $\Phi_I$ pattern.

(ii) $\Delta_A=1MHz, \Delta_B=10MHz$. In this case, $P_{LZ1}\sim 1$, $\sin\theta_1\sim 1$, $\cos\theta_1\sim 0$. The occupation probability approximately equals to:
\begin{equation}
P_1=\sin^4\theta_2+\cos^4\theta_2+2\sin^2\theta_2\cos^2\theta_2\cos \Phi_{II}.
\end{equation}
This combination reveals the main feature of $\Phi_{II}$ pattern.

(iii) $\Delta_A=10MHz, \Delta_B=1MHz$. In this case, $P_{LZ2}\sim 1$, $\sin\theta_2\sim 1$, $\cos\theta_2\sim 0$. The occupation probability approximately equals to:
\begin{eqnarray}
\lefteqn{P_1=\sin^4\theta_1+\cos^4\theta_1+{}}\nonumber\\
&&{}+2\sin^2\theta_1\cos^2\theta_1\cos (\Phi_I+\Phi_{II}).
\end{eqnarray}
This combination reveals the main feature of $\Phi_I+\Phi_{II}$ pattern.

(iv) $\Delta_A=\Delta_B=10MHz$. The occupation probability is in the form of Eq.(\ref{TwoTLSs}).

Provided that LZ transition probability takes the asymptotic form in Eq.(\ref{PLZ}), interference patterns in the 4 cases above calculated based on Eq.(\ref{TwoTLSs}) (see Fig.\ref{DiffCombination}(A)-(D)), agree well with the numerical simulations (see Fig.\ref{DiffCombination}(E)-(H)), except the slight modulation in one interference fringe involved in the numerical results. The modulation is caused by the fluctuation of actual LZ transition probabilities around its asymptotic form\cite{Guozhu}.

\begin{figure}
\includegraphics[width=2.75in]{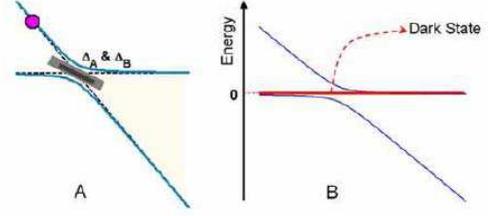}
\caption{Dark state in a hybrid qubit. (A) The two TLSs share the same location in the energy diagram. (B) Calculated energy spectrum from Hamiltonian (\ref{DarkStateH}).The red branch keeps staying at the state $|TLS1\rangle$.} \label{Darkstate}
\end{figure}

In addition, it is noteworthy that in a qubit-two-TLSs hybrid system, it is completely possible there exists a smaller anticrossing screened by a larger one. Although the LZS interference is just the same as that of a single anticrossing, one branch in the spectrum is always at the excited state of TLS, $|TLS1\rangle$ (see red line in Fig.\ref{Darkstate}(B)). The system's simplified Hamiltonian in a basis formed by {$|1g_1g_2\rangle, |0e_1g_2\rangle, |0g_1e_2\rangle$} reads,
\begin{equation}\label{DarkStateH}
\hat{H}_D=\left[\begin{array}{cccc}
\omega & \Omega_1/2 & \Omega_2/2\\
\Omega_1/2 & 0 & 0\\
\Omega_2/2 & 0 & 0
\end{array}\right],
\end{equation}
where $\omega$ is the detuning between qubit and TLS. The eigenstate corresponding to the particular branch is:
\begin{equation}
|\Phi_D\rangle=\frac{\Omega_2}{\sqrt{\Omega_1^2+\Omega_2^2}}|0e_1g_2\rangle-\frac{\Omega_1}{\sqrt{\Omega_1^2+\Omega_2^2}}|0g_1e_2\rangle.
\end{equation}
With no ingredient of $|1g_1g_2\rangle$ in it, the state $|\Phi_D\rangle$ is analogous to the Dark State in quantum optics, which has no ingredient of excited state. In this sense, we can call this ``Dark state of a hybrid qubit''. Vacancy of $|1g_1g_2\rangle$ state could largely reduce the influence of environment on the dark state, and thereby has potential use in information storage. The specific method to realize this is in need of further investigation.

\section{Fourier Transform}
Furthermore, Fourier transform is a helpful tool in identifying the individual phase components of the compound LZS interference. Supposing the two-anticrossing system follows a linear ramp traversing the anticrossing regions, FT of the occupation probability in Eq.(\ref{TwoTLSs}) is found to be,
\begin{equation}
\begin{split}
P_{FT}(k_T,A_{pulse})&=B_0+B_1\delta(k_T-k_{T1}(A_{pulse}))\\
&+B_2\delta(k_T-k_{T2}(A_{pulse}))\\
&+B_3\delta(k_T-k_{T1}(A_{pulse})-k_{T2}(A_{pulse})),
\end{split}
\end{equation}
where
\begin{equation}
\begin{split}
&B_0=\cos^4\theta_1+\sin^4\theta_1\cos^4\theta_2+\sin^4\theta_1\sin^4\theta_2,\nonumber\\
&B_1=2\sin^2\theta_1\cos^2\theta_2\cos^2\theta_1,\nonumber\\
&B_2=2\sin^4\theta_1\sin^2\theta_2\cos^2\theta_2,\nonumber\\
&B_3=2\sin^2\theta_1\sin^2\theta_2\cos^2\theta_1,\nonumber\\
&k_{T1}=\epsilon_{12}-\frac{(\epsilon_1+\epsilon_2)\epsilon_{12}}{2sA_{pulse}},\nonumber\\
&k_{T2}=\frac{(sA_{pulse}-\epsilon_{12})^2}{2sA_{pulse}}.\nonumber
\end{split}
\end{equation}
Here $s$ is the diabatic energy-level slope, and $\epsilon_{12}$ is the detuning between the location of anticrossing I and II.

Therefore, it is expected that FT reveals a highly ordered structure of one-dimensional arcs in Fourier space. (A) $\Phi_I$ pattern dominates, i.e., $\sin\theta_2\simeq0$. In this case, $B_2,B_3\simeq0$, and $P_{FT}\simeq2\sin^2\theta_1\cos^2\theta_1\delta(k_T-k_{T1}(A_{pulse}))$.
Apparently, there is only one curve corresponding to $k_T=k_{T1}$ after FT. (B) $\Phi_{II}$ pattern dominates, i.e., $\cos\theta_1\simeq0$. In this case, $B_1,B_3\simeq0$, and $P_{FT}\simeq2\sin^4\theta_1\sin^2\theta_2\cos^2\theta_2\delta(k_T-k_{T2}(A_{pulse}))$. Only $k_T=k_{T2}$ shows up after FT. (C) $\Phi_I+\Phi_{II}$ pattern dominates, i.e.,
$\cos\theta_2\simeq0$. In this case, $B_1,B_2\simeq0$, and $P_{FT}\simeq2\sin^2\theta_1\cos^2\theta_1\delta(k_T-k_{T1}(A_{pulse})-k_{T2}(A_{pulse}))$.
Only $k_T=k_{T3}$ shows up after FT. (D) All three patterns including $\Phi_I$, $\Phi_{II}$ and $\Phi_I+\Phi_{II}$ dominate, i.e., $\sin^2\theta_1\simeq\sin^2\theta_2\simeq1/2$. In this case, $B_1\simeq B_3\simeq 1/4$ and $B_2\simeq 1/8$. All three curves
corresponding to $k_T=k_{T1}$, $k_T=k_{T2}$ and $k_T=k_{T3}$ can be observed after FT.

This is explicitly demonstrated in Fig.\ref{FT}, which is based upon the Discrete Fourier Transform method:
\begin{equation}
P_{DFT}(k_T,A_{pulse})=\sum_{j=1}^N P_{1j}(T,A_{pulse})\omega_N^{(j-1)(k-1)},
\end{equation}
where $\omega_N=exp(-2\pi i/N),N=1,2,...$. As labeled in the figure, different curves correspond to different phase patterns, which is in good agreement with our analysis above.  In the multi-anticrossing system, if N anticrossings take part in the LZS interference, $C_{N+1}^2$ phase components can be observed in its FT and vice versa. Therefore, FT provides a means to ascertain how many TLSs are effectively coupled to qubit.

\section{SUMMARY}

In summary, we present a simple form of analytic expression for describing the coherent dynamics of a driven multi-anticrossing chain. The oscillatory population of the hybrid system remained at the initial state exhibits a rich pattern of LZS interference in the two-dimensional phase space parameterized by pulse width and driving amplitude. In the N-anticrossing chain, the resulted compound interference is the addition of $C_{N+1}^2$ patterns governed by two transmitted paths, whose weights rely on the LZ transition amplitude.
This is clearly demonstrated by their Fourier transforms of the pulse width, which serve as a useful tool in offering information on energy spectrum. Although the intrinsically random nature of TLSs precludes the direct control of their distribution and coupling strength with qubit, our discussion of possible types of special hybrid qubit can be used to understand some observed patterns and to predict future experimental phenomena, in particular considering the rapid technological advancement of a macroscopic device with an atomic-sized system\cite{Matthew}. Moreover, it is straightforward to apply our method to a superconducting flux qubit, where resonant tunneling between its double well potential forms a multi-anticrossing net. Although effort has been devoted to study that system\cite{Xushuang}, our universal model can give a more general description of the system.

\section{ACKNOWLEDGMENTS}

This work was supported in part by the State Key Program for Basic Research of China (Grant No. 2006CB921801), and NSFC (National Natural Science Foundation of China, Grant No. 10725415) .

\end{document}